\documentclass[10pt,prb,aps,floatfix,superscriptaddress,showpacs,footinbib,twocolumn,longbibliography]{revtex4-2}

\usepackage[colorlinks=true,urlcolor=blue,citecolor=blue,linkcolor=blue]{hyperref}
\usepackage{graphicx,mwe} 
\usepackage{xcolor}
\usepackage{bm}
\usepackage{bbm}
\usepackage{ulem}
\usepackage{braket}
\usepackage[utf8]{inputenc}
\usepackage[T1]{fontenc}
\usepackage{amsmath,mathtools}
\usepackage{mathrsfs}
\usepackage{booktabs}       
\usepackage{longtable}
\usepackage{amssymb}
\usepackage{soul}
\usepackage{float}
\usepackage{hyperref}
\def\umphys{
    Department of Physics, University of Michigan,
    Ann Arbor, Michigan 48109, USA
}
\def\uwphys{
Institute of Theoretical Physics, Faculty of Physics, University of Warsaw, Warsaw, Poland
}
\def\ustcinstitute{
Suzhou Institute for Advanced Research, University of Science and Technology of China, Suzhou 215123, People’s Republic of China
}
\def\ustcschool{
School of Artificial Intelligence and Data Science, University of Science and Technology of China, Suzhou 215123, People’s Republic of China
}
\def\cas{
Beijing National Laboratory for Condensed Matter Physics and Institute of Physics, \\Chinese Academy of Sciences, Beijing 100190, China
}

\begin{document}
\title{Multi-orbital dynamical mean-field theory with a complex-time solver
}
\author{Yang Yu}
\affiliation{\umphys}
\author{Lei Zhang}
\affiliation{\umphys}
\author{Emanuel Gull}
\affiliation{\umphys}
\affiliation{\uwphys}
\author{Xiaodong Cao}
\email{xdcao@ustc.edu.cn}
\affiliation{\ustcinstitute}
\affiliation{\ustcschool}
\author{Xinyang Dong}
\email{dongxy@iphy.ac.cn}
\affiliation{\cas}

\date{\today}

\begin{abstract}
We present the combination of a complex-time tensor-network impurity solver with an analytic continuation scheme based on exponential fitting as an efficient framework for single and multi-orbital dynamical mean-field  calculations.
By performing time-evolution along a complex-time contour, the approach balances computational cost with the difficulty of spectral recovery, offering greater flexibility than methods confined to the real or imaginary axis. By complementing the complex-time evolution with an exponential fitting scheme, we faithfully extract real-time information at negligible cost. The resulting method obtains high-resolution spectra at a significantly lower computational cost than real-time evolution, offering a promising tool for {\it ab initio} studies of strongly correlated materials.
\end{abstract}

\maketitle

\section{Introduction}

Dynamical mean-field theory (DMFT) has become a powerful and widely used framework for studying strongly correlated electron systems. In its single-orbital formulation, DMFT captures local electronic correlations and successfully describes phenomena such as the Mott metal–insulator transition in the infinite coordinate number limit \cite{Metzner_dmft_1989, Georges_dmft_1996}. Multi-orbital DMFT extends this capability to more realistic materials by including orbital degrees of freedom, allowing one to address local orbital dependent physics such as Hund’s coupling effects \cite{Kotliar_dmft_2006}. 
In DMFT, the lattice problem is mapped onto an effective quantum impurity model, which is simpler than the original lattice model yet remains a highly nontrivial quantum many-body problem 
whose solution in general requires non-perturbative numerical methods. These methods are known as the impurity solvers.
A variety of such solvers have been developed, including self-consistent diagrammatic approaches \cite{Pruschke_ipt_1990,Keiter_nca_1970,Pruschke_oca_1989,Kotliar_dmft_2006}, continuous-time quantum Monte Carlo (CT-QMC) \cite{Rubtsov_ctint_2005,Werner_cthyb_2006,Gull_ctaux_2008,Gull_ctqmc_2011,Cohen_inchworm_2013}, exact diagonalization (ED) \cite{Caffarel_ED_1994, Capone_ED_2007, Iskakov_ED_2018} and quantum chemistry inspired methods \cite{Zigid_ci_2012,Shee_ccsd_2019,Zhu_ccsd_2019},  numerical renormalization group (NRG) methods \cite{Wilson_NRG_1975,Pruschke_NRG_2000,Bulla_NRG_2008}, and tensor-network-based methods \cite{White_DMRG_1992,White_DMRG_1993, Schollwock_DMRGReview_2011,Orus_TNSReview_2014,Wolf_mps_2015,Yuriel_tensor_2022,Kim_tci_2025,Matsuura_tci_2025,Geng_tci_2025}. 

Within DMFT, the central quantity of interest is the single-particle spectral function, which can be directly compared to spectroscopic experiments. 
Obtaining accurate spectral functions presents a significant computational challenge. Methods that operate in imaginary time or imaginary frequency require analytical continuation \cite{Jarrell_maxent_1996,Sandvik_SAC_1998,Vidberg_pade_1977,Fei_Nev_2021,Huang23,Ying_pole_2022a,Ying_pole_2022b,Zhang_minipole_2024_a} to real frequencies, which often introduces large uncertainties, especially in the high-energy features of the spectrum. 
Real-frequency or real-time solvers such as NRG or ED can circumvent this difficulty, but the exponential scaling of the Hilbert space restricts them to small or highly symmetric impurity problems.
Among emerging techniques, tensor-network-based methods are promising: they are essentially noise-free, do not suffer from exponential scaling in typical application, and treat both low- and high-energy scales on an equal footing.
They can resolve fine spectral features across the metal-insulator transition \cite{Ganahl_mps_2015} and provide accurate impurity spectral functions for embedding calculations of real materials \cite{Bauernfeind_ftps_2017, cao_tree_2021} within the zero-temperature real-frequency DMFT framework.
However, capturing long-time correlations to resolve sharp low-energy features remains a significant challenge. The rapid growth of entanglement with time demands large bond dimensions, making calculations for complex multi-orbital systems especially demanding. 

Complex-time tensor network impurity solvers have been proposed to address this problem \cite{cao_complextime_2024, grundner_complextime_2024}. By evolving the system along a properly chosen contour in the complex-time plane, these approaches suppress entanglement growth, thereby significantly reducing the required bond dimension and accelerating simulations.
Once the complex-time solution has been obtained, the real-frequency spectral function can be constructed from the numerical data on the complex-time contour.
Methods such as Taylor expansion \cite{cao_complextime_2024}, parallel-contour evaluation \cite{grundner_complextime_2024}, maximum entropy (MaxEnt) \cite{Jarrell_maxent_1996,grundner_complextime_2024}, and stochastic analytic continuation (SAC) \cite{Sandvik_SAC_1998,wang_mettscomplex_2025} have been used for this purpose.
Since this step is required at every iteration of real-frequency DMFT to solve the Dyson equation and update the impurity model, it is essential that it be robust, efficient and accurate.
Existing approaches either require additional evaluations along complex contours, which increases the computational cost, or they suffer from limited accuracy~\cite{cao_complextime_2024,grundner_complextime_2024,wang_mettscomplex_2025}.
Developing a robust and efficient method for extracting real-frequency spectra from complex-time data therefore remains an important research direction for enabling reliable studies of  correlated systems within this framework. 

In this manuscript, we propose an exponential fitting approach to reconstruct real-time data from simulated complex-time results, enabling direct extraction of real-frequency information from the fitted coefficients and exponents.
The algorithm incurs negligible computational cost and depends on a single parameter controls the fitting error level.
The complex-time simulations are controlled by an angle $\alpha$ in the complex plane, with $\alpha=0$ the real-time simulation and $\alpha \rightarrow \pi/2$ approaching imaginary-time evaluation.
We demonstrate that our method reliably recovers the spectrum at low to intermediate angles, and is stable for both single- and multi-orbital DMFT simulations.

The reminder of this paper is organized as follows:
In Section~\ref{sec:method}, we outline the real-frequency DMFT framework as well as the complex-time formalism, and introduce our proposed analytic continuation method based on exponential fitting.
Section~\ref{sec:results} presents benchmark calculations, beginning with the single-orbital Anderson impurity model and extending to single- and multi-orbital DMFT. 
Section~\ref{sec:conclusion} summarizes our key results and discusses their broader context and potential applications.

\section{Methods} \label{sec:method}

The primary focus of this manuscript is on zero-temperature DMFT calculations employing a complex-time tensor network impurity solver \cite{cao_complextime_2024,grundner_complextime_2024}. In this section, we first briefly review the DMFT formalism. Next, we define the complex-time Green’s functions. Finally, we present the main methodological contribution of this work: using an exponential fitting method to accurately and efficiently reconstruct the real-frequency spectral function from complex-time Green’s functions.

\subsection{DMFT} \label{subsec:complex_time_DMFT}

In this work, we focus for simplicity on computing the spectral function of models on Bethe lattice in the infinite-coordination limit within zero-temperature real-frequency DMFT.
The corresponding Hamiltonian of the impurity problem is $\hat{H} = \hat{H}_\text{loc} + \hat{H}_\text{bath} + \hat{H}_\text{hyb}$, with
\begin{align}
    \hat{H}_\text{loc} &= \sum_{\nu} \varepsilon_{\nu_{1}\nu_{2}}\hat{d}_{\nu_{1}}^\dagger \hat{d}_{\nu_{2}} + \hat{H}_\text{int},
    \\
    \hat{H}_\text{int}
    &=
    U \sum_i \hat n_{i\uparrow}\hat n_{i\downarrow}
    + \sum\limits_{i<j,\sigma\sigma^{\prime}}\left(U^{\prime}-J \delta_{\sigma \sigma^{\prime}}\right) \hat{n}_{i \sigma} \hat{n}_{j \sigma^{\prime}}
    \nonumber\\
    &- J \sum_{i\neq j}
    \hat d^\dagger_{i\uparrow}\hat d_{i\downarrow}
    \hat d^\dagger_{j\downarrow}\hat d_{j\uparrow}
    + J \sum_{i\neq j}
    \hat d^\dagger_{i\uparrow}\hat d^\dagger_{i\downarrow}
    \hat d_{j\downarrow}\hat d_{j\uparrow},
    \\
    \hat{H}_\text{bath} &= \sum_{b=1}^{N_b}
    \sum_{\kappa}\epsilon^b_{\kappa}\hat{c}_{b\kappa}^\dagger\hat{c}_{b\kappa},
    \\
    \hat{H}_\text{hyb} &= 
     \sum_{b=1}^{N_b}\sum_{\kappa \nu} 
   V^b_{\nu \kappa}\hat{d}_{\nu}^\dagger \hat{c}_{b\kappa} + {h.c.} ,
\end{align}
where $\hat{d}_{i\sigma}$ and $\hat{d}^{\dagger}_{j\sigma}$ are the fermion annihilation and creation operators for impurity orbitals $i, j$ of spin $\sigma\in \set{\uparrow,\downarrow}$, $\nu$ and $\kappa$ are the combined spin-orbital indices for impurity and bath, and $b$ is bath site index.
$U$ is the intra-orbital Hubbard interaction, $J$ is the Hund's coupling, and $U' = U-2J$ for a rotationally invariant system. 
The impurity is coupled to the bath via $\hat{H}_\text{hyb}$, with the parameters $V^b_{\nu \kappa}$ and $\epsilon^{b}_{\kappa}$ determined from the hybridization function
\begin{align}
    \Delta_{\nu_{1}\nu_{2}}(\omega) = \sum_{b=1}^{N_b} \sum_{\kappa} \frac{V^{b}_{\nu_{1}\kappa} V^{b*}_{\nu_{2}\kappa}}{\omega + i 0^+ - \epsilon^{b}_{\kappa}}.
\end{align}
We employ a spin-symmetric bath and restrict to the normal (paramagnetic) state, and therefore omit spin indices for all calculated quantities.

At the DMFT self-consistency, the impurity Green's function equals the local lattice Green's function \cite{Georges_dmft_1996}. 
The self-consistency is realized by starting from an initial guess (such as the non-interacting solution) and iterating until convergence is reached. 
Within each iteration, the hybridization function is updated with the impurity spectral function
\begin{align}
    -\frac{1}{\pi} \text{Im} \Delta_{ij}(\omega) = \frac{D^2}{4} A_{ij}(\omega) ,
    \label{eq:hybridization}
\end{align}
where $D$ is the half bandwidth of the initial semielliptic noninteracting density of states $A^{(0)}(\omega)=\frac{2}{\pi D} [1-\left( \frac{\omega}{D}\right)^2 ]^{1/2}$.  $A_{ij}(\omega) = -\frac{1}{\pi} \text{Im} G^{\mathrm{R}}_{ij}(\omega)$ is the impurity spectral function, with $G^{\mathrm{R}}_{ij}$ the retarded impurity Green's function.

\subsection{Complex-time Green's functions} \label{subsec:complex_time_GF}

\begin{figure}[tb]
    \centering
    \includegraphics[width=0.7\linewidth]{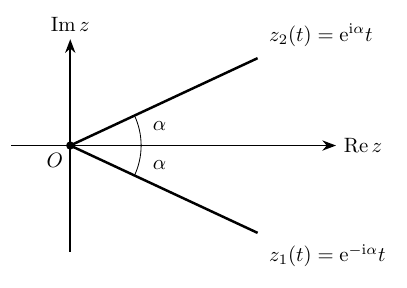}
    \caption{Fixed-angle contours parameterized by $t\geq 0$ and $\alpha \in [0,\pi/2]$.}
    \label{fig:complex_time_contour}
\end{figure}

Within tensor-network methods, a common approach to obtain the real-frequency retarded impurity Green’s function is to compute the real-time Green’s function, and then perform a direct Fourier transform \cite{Ganahl_mps_2015}.
However, real-time evolution within tensor network solvers is limited by entanglement growth: the required bond dimension increases with simulation time, making long time calculations expensive \cite{cao_complextime_2024, grundner_complextime_2024}.
To overcome this difficulty, Refs.~\onlinecite{cao_complextime_2024, grundner_complextime_2024} developed a complex-time impurity solver that performs time evolution along a contour $z(t)$ in the complex-time plane.
The fermionic Green's functions on this contour are defined as \cite{cao_complextime_2024, grundner_complextime_2024}
\begin{subequations} \label{eqn:general_complex}
\begin{align}
    G_{\nu_{1}\nu_{2}}^{>}(z_{1}) 
    &\coloneqq - \mathrm{i} \braket{g|  \hat{c}_{\nu_{1}} \mathrm{e}^{- \mathrm{i} (\hat{H}-E_{g}\hat{I}) z_1(t) } \hat{c}^{\dagger}_{\nu_{2}}|g}, \label{eqn:general_greater}
    \\
    G_{\nu_{1}\nu_{2}}^{<}(z_{2}) 
    &\coloneqq   \mathrm{i} \braket{g| \hat{c}^{\dagger}_{\nu_{2}} \mathrm{e}^{\mathrm{i} (\hat{H}-E_{g}\hat{I}) z_2(t) } \hat{c}_{\nu_{1}}|g}, \label{eqn:general_lesser}
\end{align}
\end{subequations}
where a unique ground state $\ket{g}$ is assumed and the corresponding energy $E_{g}$ is explicitly subtracted from the Hamiltonian $\hat{H}$.
The real-time greater $G_{\nu_{1}\nu_{2}}^{>}(t)$ and lesser $G_{\nu_{1}\nu_{2}}^{<}(t)$ Green's functions can be treated as a special case of the complex-time Green's functions with $z_1(t) = z_2(t) = t$.
We focus on the complex-time Green's functions $G_{\nu_{1}\nu_{2}}^{\gtrless}(t;\alpha)$ defined on a specific contour $z_1(t) = \mathrm{e}^{-\mathrm{i}\alpha}t$, $z_2(t) = \mathrm{e}^{\mathrm{i}\alpha}t$ as shown in Fig.~\ref{fig:complex_time_contour}. To ensure their well-behaved long tails, we restrict the parameters to $t\geq 0$ and $\alpha \in [0,\pi/2]$. These complex-time Green's functions are related to their real-time counterparts through
\begin{subequations} \label{eqn:complex_relation}
\begin{align}
   &G_{\nu_{1}\nu_{2}}^{>}(t;\alpha) = G_{\nu_{1}\nu_{2}}^{>}(t\to \mathrm{e}^{-\mathrm{i}\alpha}t), \label{eqn:greater_relation}
   \\
   &G_{\nu_{1}\nu_{2}}^{<}(t;\alpha) = G_{\nu_{1}\nu_{2}}^{<}(t\to \mathrm{e}^{\mathrm{i}\alpha}t). \label{eqn:lesser_relation}
\end{align}
\end{subequations}

\subsection{Spectrum reconstruction via exponential fitting}\label{subsec:exponential_fitting}

The DMFT self-consistency loop requires reconstructing the retarded impurity Green's function
$G^{\mathrm{R}}_{ij}(\omega)$ on the real-frequency axis [cf. Eq.~\eqref{eq:hybridization}] from the complex-time greater and lesser Green's functions $G^{\gtrless}_{ij}(t;\alpha)$.
This procedure is analogous to the conventional analytical continuation problem from the imaginary-time domain ($\alpha=\pi/2$ in Fig.\ref{fig:complex_time_contour}) to the real-frequency domain, which is notoriously ill-conditioned. 
Several ways have been proposed for this purpose in previous works \cite{cao_complextime_2024, grundner_complextime_2024, wang_mettscomplex_2025}.
However, these approaches either rely on extra computationally expensive evaluations or lack sufficient accuracy, particularly for large values of the complex angle $\alpha$ (see Appendix~\ref{app:maxent} for a MaxEnt example).

Here we show that these limitations can be overcome using an exponential fitting method \cite{Zhang_minipole_spectrum_2025, Erpenbeck_fitting_2025}.
The basic assumption is that each matrix element of the complex-time Green's functions is well approximated by a sum of complex exponentials
\begin{align}
    G^{>}_{ij}(t;\alpha) \approx \sum_{l=1}^{M^{>}} R_{l}^{>} \mathrm{e}^{s_{l}^{>}t}, 
    \quad
    G^{<}_{ij}(t;\alpha) \approx \sum_{l=1}^{M^{<}} R_{l}^{<} \mathrm{e}^{s_{l}^{<}t}. \label{eqn:fitting_complex}
\end{align}
The real-time Green's functions are then given by [cf. Eq.~\eqref{eqn:complex_relation}]
\begin{align}\label{eqn:obtain_real_time}
    G^{>}_{ij}(t) \approx \sum_{l=1}^{M^{>}} R_{l}^{>} \mathrm{e}^{s_{l}^{>} \mathrm{e}^{\mathrm{i}\alpha}t},
\, \,
    G^{<}_{ij}(t) \approx \sum_{l=1}^{M^{<}} R_{l}^{<} \mathrm{e}^{s_{l}^{<} \mathrm{e}^{-\mathrm{i}\alpha}t},
\end{align}
which can be used to compute the retarded Green's function
\begin{align}
    G^{\mathrm{R}}_{ij}(t) &= \theta(t) [G^{>}_{ij}(t)-G^{<}_{ij}(t)] , \label{eqn:construct_retarded}
\\
    G^{\mathrm{R}}_{ij}(\omega) &= \int_{-\infty}^\infty \mathrm{d}t\, \mathrm{e}^{\mathrm{i}\omega t} G_{ij}^{\mathrm{R}}(t) . \label{eqn:gf_ft}
\end{align}
Determining the parameters $\{ M^{\gtrless}, R_{l}^{\gtrless}, s_{l}^{\gtrless} \}$ in Eq.~\ref{eqn:fitting_complex} is a standard signal processing problem. A variety of numerical methods exist for this purpose \cite{Prony_Prony_1795,Roy_ESPRIT_1989,Hua_MP_1990,Sarkar_MP_1995,Schmidt_MUSIC_1986,Schmid_DMD_2010,Potts_ESPRIT_2013}.
We employ the Estimation of Signal Parameters via Rotational Invariance Techniques (ESPRIT) \cite{Roy_ESPRIT_1989,Potts_ESPRIT_2013} due to its superior performance in approximating the real-time Green's function \cite{Zhang_minipole_spectrum_2025,Erpenbeck_fitting_2025}.
We briefly summarize the algorithm below, with details available in the original literature \cite{Roy_ESPRIT_1989,Zhang_minipole_2024_b}.

To approximate a function $f(t) \approx \sum_{l=1}^{M} R_{l} \mathrm{e}^{s_{l}t}$ sampled on an equidistant time grid $\{ t_i = i \Delta t | i =0,1,\cdots,N-1\}$ within a tolerance $\epsilon$ using the ESPRIT algorithm, we first form the Hankel matrix $H$ from the given data and compute its singular value decomposition
\begin{align}
    H =
    \begin{pmatrix}
    f(t_0)        & f(t_1)        & \cdots & f(t_L)        \\
    f(t_1)        & f(t_2)        & \cdots & f(t_{L+1})    \\
    \vdots        & \vdots        & \ddots & \vdots        \\
    f(t_{N-L-1})  & f(t_{N-L})    & \cdots & f(t_{N-1})
    \end{pmatrix}
    = U \Sigma W \, ,
\end{align}
in which $N$ is the number of data points, and $L$ is chosen as $2N/5$~\cite{Sarkar_MP_1995}.
The number of exponents $M$ is determined as the smallest index that satisfies $\sigma_{M} < \epsilon$, where $\{\sigma_i\}$ are the singular values of $H$. 
The exponents $\{s_l\}$ are then extracted from the eigenvalues $\{x_l = \mathrm{e}^{s_l\Delta t}\}$ of the matrix $F = (W_0^{T})^{+} W_1^{T}$, with $W_s = W(0:M-1,\, s:L+s-1), s = 0,1$.
Finally, the corresponding weights $\{R_l\}$ are obtained by solving an overdetermined Vandermonde system
\begin{align}
    \begin{pmatrix}
    f(t_0) \\
    f(t_1) \\
    \vdots \\
    f(t_{N-1})
    \end{pmatrix}
    =
    \begin{pmatrix}
    1 & 1 & \cdots & 1 \\
    x_1 & x_2 & \cdots & x_M \\
    \vdots & \vdots & \ddots & \vdots \\
    x_1^{N-1} & x_2^{N-1} & \cdots & x_M^{N-1}
    \end{pmatrix}
    \begin{pmatrix}
    R_1 \\
    R_2 \\
    \vdots \\
    R_M
    \end{pmatrix}.
\end{align}
This procedure yields an approximation of the form of Eq.~\ref{eqn:fitting_complex}.

Directly performing the numerical Fourier transform in Eq.~\ref{eqn:gf_ft} requires long simulation times, small time discretization, and artificial broadening to suppress spurious features, particularly when the Green’s function decays slowly.
We therefore employ a complex pole representation of the retarded Green's function as described in Refs.~\onlinecite{Zhang_minipole_2024_a, Zhang_minipole_2024_b}
\begin{align}
    G^{\mathrm{R}}_{ij}(\omega) \approx \sum_{l=1}^M \frac{A_{l}}{\omega + \mathrm{i}0^{+} - \xi_{l}},
\end{align}
where 
$\{ \xi_l \}$ are $M$ complex poles in the lower half of the complex frequency plane with corresponding complex weights $\{ A_l \}$.
As shown in Ref.~\onlinecite{Zhang_minipole_spectrum_2025}, the residue theorem ensures these poles and weights can be determined from the real-time Green's functions
\begin{align}
    G^{\mathrm{R}}_{ij}(t) \approx -\mathrm{i}\theta(t)\sum_{l=1}^{M} A_{l} \mathrm{e}^{-\mathrm{i}\xi_{l} t}. \label{eqn:retarded_time_fitting}
\end{align}
The parameters $\{A_l, \xi_l\}$ in this expression are again extracted using the ESPRIT algorithm after $G^{\mathrm{R}}_{ij}(t)$ is constructed following Eqs.~\eqref{eqn:obtain_real_time} and~\eqref{eqn:construct_retarded}. 
See Ref.~\onlinecite{Zhang_minipole_2024_a, Zhang_minipole_spectrum_2025} for more discussions regarding ESPRIT and the complex pole representation of the Green's function.

In practice, to ensure a decaying time series, when using ESPRIT to fit $G^{\gtrless}_{ij}(t;\alpha)$,
we discard unphysical exponents that would cause divergence after rotating back to the real-time axis, i.e., modes with
\(|e^{s_l^{>} e^{i\alpha} \Delta t}|>1\) or \(|e^{s_l^{<} e^{-i\alpha} \Delta t}|>1\)
before performing the least-square fit to find the weights $R_{l}^{\gtrless}$.
To ensure a smooth spectral function and eliminate unphysical features, an additional filtering step can be applied during the final ESPRIT fit in Eq.~\eqref{eqn:retarded_time_fitting}, retaining only poles with real part smaller than a chosen bound before performing the least-square fit for the amplitudes $A_l$. 
As shown in Refs.~\onlinecite{Zhang_minipole_2024_b,Erpenbeck_fitting_2025}, these types of filtering are useful to remove unphysical overfitting of the data and work well in practice.

We note that the computational cost of the ESPRIT fit is negligible in our applications, and the ESPRIT-based analytic continuation procedure requires no additional evaluations on the complex-time contour. 
This efficiency makes it particularly useful for simulation frameworks that require repeated calculations such as DMFT.
For all calculations in this manuscript, we consider systems with a diagonal hybridization function and half bandwidth $D=2$ unless otherwise stated. Consequently, the fitting and Fourier transform are performed separately for each diagonal component of the matrix-valued Green’s function. 
For the more general case of multi-orbital systems with off-diagonal hybridization, a more physically insightful approach is to use a set of shared complex poles with matrix-valued weights, which can be implemented via the matrix-valued ESPRIT fitting and complex-pole representation formalism developed in Ref.~\cite{Zhang_minipole_2024_b}.

\section{Results} \label{sec:results}

For all results presented in this section, the real- and complex-time impurity Green's functions are computed using a tree tensor network (TTN) impurity solver \cite{Shi_tree_2006, Murg_tree_2010, cao_tree_2021}. 
The time evolution in Eq.~\eqref{eqn:general_complex} is performed using a hybrid scheme: physical tensors are updated with two-site time-dependent variational principle (TDVP) \cite{Haegeman_tdvp_2011, Haegeman_tdvp_2013, Haegeman_tdvp_2016, Vanderstraeten_tdvp_2019}, while auxiliary tensors are updated primarily with single-site TDVP, supplemented by a two-site TDVP update every 10 time steps (see Ref.~\onlinecite{cao_tree_2021} for details).
The time interval is chosen to be $D \Delta t = 0.1$, which proves sufficient for the simulations in this work. 
All calculations are performed in the ``natural orbital'' basis to reduce computational costs \cite{Lu_no_2014, lu_no_2019}.
The error level $\epsilon$ is set to $10^{-6}$ when we use ESPRIT to fit the complex-time series, which enables us to obtain accurate spectral functions without overfitting the input data.

\subsection{Single-orbital Anderson impurity model}
\begin{figure}[tb]
    \centering
    \includegraphics[scale=1]{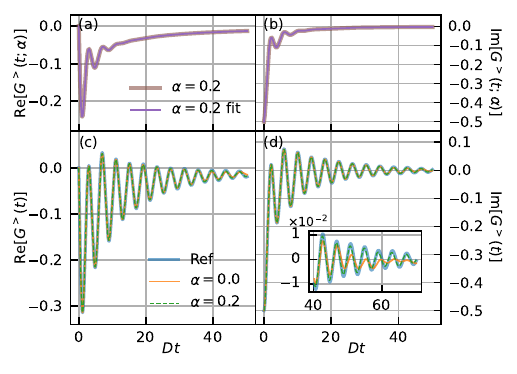}
    \caption{Complex and real-time evolution results at $U=2D$ and $N_b=199$. Panels (a) and (b): Real and imaginary parts of the ESPRIT fitted complex-time Green's function with $\alpha=0.2$. Panels (c) and (d): Comparison of the real and imaginary parts of the real-time Green’s function obtained by direct real-time evolution and by analytic continuation from complex-time data. The reference curve is computed by real-time evolution with $\chi=1500$, all other results use $\chi=80$.}
    \label{fig:recover}
\end{figure}

We first benchmark the analytic continuation of complex-time Green's function using the single-orbital Anderson impurity model (SAIM) at half-filling, following the examples in Refs.~\onlinecite{cao_complextime_2024, grundner_complextime_2024}. 
Fig.~\ref{fig:recover} shows the real-time series and the $\alpha=0.2$ complex-time series computed at $U = 2D$ and $N_b = 199$ bath sites. 
The results in panels (a) and (b) show that ESPRIT accurately reproduces the computed complex-time series. 
Panels (c) and (d) then compare the real-time Green's function obtained by direct real-time evolution and analytical continuation of complex-time data, both computed with bond dimension $\chi=80$, against a high-accuracy real-time reference computed with $\chi=1500$. 
While both approaches agree well with the reference at early times, the complex-time results exhibit substantially smaller deviations at longer times (see inset), demonstrating that complex-time simulations can capture accurate long-time dynamics with a smaller bond dimension.

\begin{figure}[tbh]
    \centering
    \includegraphics[scale=1]{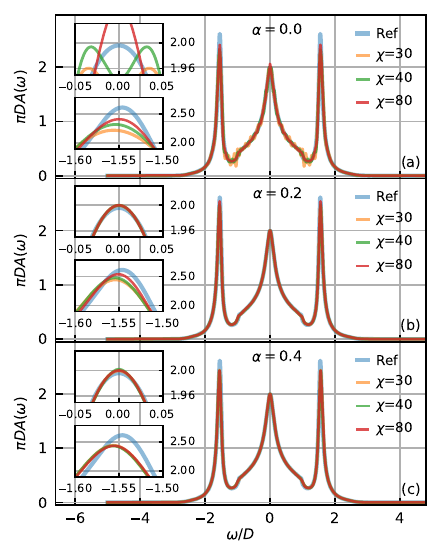}
    \caption{Comparison of spectral functions obtained from different complex angles at $U=2D$, $N_b=199$ and $D t_\text{max}=80$. (a) $\alpha=0.0$, (b) $\alpha=0.2$, (c) $\alpha=0.4$. The $\alpha=0.0$ results are computed by direct Fourier transform, while the others are obtained by ESPRIT fitting. The reference curve is the direct Fourier transform results of $\alpha=0.0$, $N_b=199$, $D t_\text{max}=100$ and $\chi=1500$. }
    \label{fig:spectral_imp}
\end{figure}

Fig.~\ref{fig:spectral_imp} provides a further benchmark of the complex-time method by presenting the spectral function at $U=2D$ obtained with different complex angles $\alpha = 0, 0.2, 0.4$ and final time $Dt_\text{max} = 80$.
By comparing to the reference results computed via real-time evolution with bond dimension $\chi=1500$, it is clear that the low-energy part of the spectral function, which should fulfill $\pi D A(\omega=0) = D$ according to the Friedel sum rule~\cite{Luttinger_1960,Luttinger_1961}, is well recovered by complex-time simulations at both $\alpha=0.2$ and $0.4$ using relatively low bond dimension, whereas real-time evolution requires a much larger bond dimension to converge (see insets centered at $\omega/D=0$).
For the high-energy Hubbard band structures, the recovered spectrum shows increasing deviations as the angle $\alpha$ grows (see insets centered at $\omega/D=-1.55$).
This trend is due to the progressive damping of high-frequency information as the contour deviates from the real axis, as analyzed in detail in Appendix~\ref{app:nonint}.
\begin{figure}[tbh]
    \centering
    \includegraphics[scale=0.98]{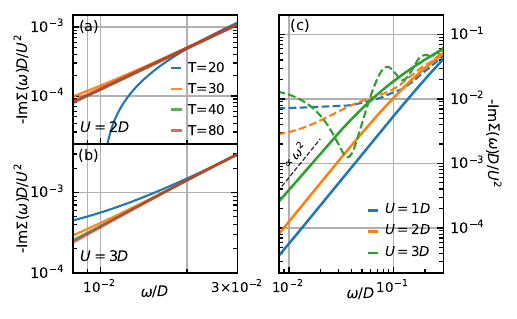}
    \caption{Self-energies computed with $N_b=399$ and $\chi=80$. Panels (a) and (b): Comparison of results extracted from complex-time evolution with $\alpha=0.4$ and different end times at $U=2D$ and $3D$. Panel (c): Comparison of results obtained from real (dashed lines) and complex (solid lines) time evolutions with $\alpha=0.4$. }
    \label{fig:sigma_imp}
\end{figure}

We show the low-energy part of the impurity self-energy $\Sigma^{R}(\omega) = \omega + \mu - \Delta^{R}(\omega) - \left[G^{R}(\omega)\right]^{-1}$ for various $U$ in Fig.~\ref{fig:sigma_imp} to demonstrate the accuracy of our method, since recovering the self-energy accurately is computationally more challenging.
Convergence tests in panels (a) and (b) indicate that using ESPRIT to extract modes from complex-time data and perform Fourier transform using the complex pole representation at $U=2D$ and $3D$ yields essentially converged results with $Dt_\text{max} = T=40$ for $\alpha=0.4$.
As shown in panel (c), we are able to recover the low-energy Fermi-liquid behavior $-\mathrm{Im}\Sigma(\omega)\propto \omega^{2}$  with precision comparable to Refs.~\onlinecite{cao_complextime_2024, grundner_complextime_2024}. 
Crucially, we achieve this by simulating at a relatively large complex angle $\alpha=0.4$, a regime challenging for previous methods ~\cite{cao_complextime_2024,grundner_complextime_2024,wang_mettscomplex_2025} (also see Appendix~\ref{app:maxent}), while avoiding the extra computational cost associated with Taylor expansions~\cite{cao_complextime_2024} or evaluations along multiple complex contours~\cite{grundner_complextime_2024}. 
Since simulations at larger complex angle $\alpha$ are faster and require smaller bond dimensions than those at smaller $\alpha$ \cite{cao_complextime_2024, grundner_complextime_2024}, this method provides a more efficient route for using the complex-time solver.

\subsection{Single-orbital dynamical mean field theory }

\begin{figure}[tbh]
    \centering
    \includegraphics[scale=1]{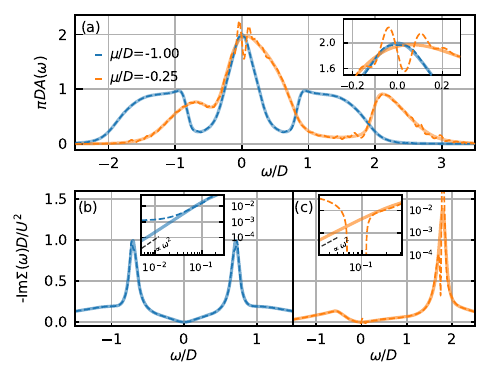}
    \caption{Comparison of real (dashed lines) and complex (solid lines) time single-orbital DMFT simulation results at $U=2D$, $N_b=179$, and $D t_\text{max}=60$. Results are shown for $\mu/D =-1.00$ ($n=1$, blue lines) with $\chi=30$ and $\mu/D=-0.25$ ($n \approx 0.72$, orange lines) with $\chi=40$. Complex-time simulations are performed with $\alpha=0.2$. Panel (a): Spectral functions. Panels (b) and (c): Imaginary part of the self-energy.}
    \label{fig:spectral_dmft}
\end{figure}
\begin{figure}[tbh]
    \centering
    \includegraphics[scale=1]{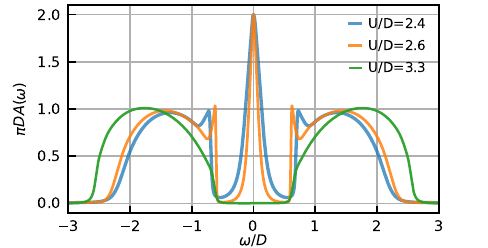}
    \caption{Single-orbital DMFT spectral functions at various $U$ computed from complex-time simulations with $\alpha=0.2$, $\chi=60$. $U/D=2.4, 2.6$: $N_b = 179$, $Dt_\text{max}=60$; $U/D=3.3$: $N_b=59$, $Dt_\text{max}=30$.}
    \label{fig:dmft_U}
\end{figure}

We now proceed to single-orbital DMFT to evaluate the stability of our method within the self-consistent loop. Fig.~\ref{fig:spectral_dmft} shows the converged spectral functions (top panels) and self-energies (bottom panels) at $U=2D$ for the half-filled (blue lines) and hole-doped (orange lines) systems. To ensure accuracy across both low- and high-energy regimes, we employ complex-time simulations with $\alpha=0.2$. For comparison, the real-time results ($\alpha=0$, dashed lines) are obtained in the final DMFT iteration using the same input hybridization function. 
The DMFT loops show stable convergence with the complex-time impurity solver combined with ESPRIT-based analytic continuation.
Moreover, as shown in the figure, the results obtained from complex-time simulations are consistently more accurate than those from real-time simulations at fixed bond dimension. This is particularly evident in the hole-doped case, where the real-time simulation fails to produce reasonable spectral functions or self-energies at the given bond dimension, instead generating unphysical structures.

We present converged DMFT results for various interaction strengths $U$ in Fig.~\ref{fig:dmft_U} to further demonstrate the capability of our method across the metal-to-insulator transition (MIT). 
In single-orbital DMFT, the MIT is a first-order transition featuring a coexistence regime $U_{c1}/D \leq U/D \leq U_{c2}/D$, with $U_{c1} \approx 2.38$ and $U_{c2} \approx 3.0$~\cite{Karski_mit_2008, Ganahl_mps_2015}. Starting from a semielliptic noninteracting density of states, we obtain the metallic solution within this coexistence regime. As shown in Fig.~\ref{fig:dmft_U}, increasing the interaction strength gradually drives the system toward an insulating solution. 
Throughout this transition, our method is able to resolve the sharp features at the inner edges of the Hubbard bands \cite{Karski_mitspec_2005, Ganahl_mps_2015}. The outer edge of the Hubbard bands in the insulating regime is nearly a step function, which is difficult for the complex pole representation to describe~\cite{Zhang_minipole_2024_a}. We therefore apply a broadening of $\eta=0.05D$ in this frequency region to suppress unphysical artifacts. 

\subsection{Multi-orbital dynamical mean field theory}

\begin{figure}[tb]
    \centering
    \includegraphics[scale=1]{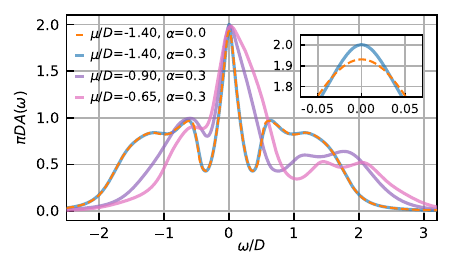}
    \caption{DMFT spectral functions of the two-orbital Hubbard model at $U = 1.6D$, $J = 0.25U$ and various doping levels. $\mu/D=-1.40$ ($n=2$): $N_b=99$, $Dt_\text{max}=50$; $\mu/D=-0.90$ ($n \approx 1.7$): $N_b=149$, $Dt_\text{max}=40$; $\mu/D=-0.65$ ($n \approx 1.5$): $N_b=279$, $Dt_\text{max}=40$. Solid lines correspond to complex-time simulations with $\alpha=0.3$, while dashed line shows the real-time result ($\alpha=0$) obtained in the final DMFT iteration by applying the same input hybridization function used in the complex-time simulation. The bond dimension is $\chi=40$ in all cases.}
    \label{fig:2orb_dmft}
\end{figure}

A reliable real-frequency impurity solver is particularly useful for multi-orbital systems, as it is a key component of real material embedding methods \cite{Kotliar_dmft_2006, Zgid_seet_2017}. 
However, obtaining these results via direct real-time evolution is computationally expensive, and complex-time simulations offer a promising way to improve efficiency in this regime. 
To demonstrate the applicability of our approach in this context, we present results for two- and three-orbital DMFT calculations in this section.

We first benchmark our method using a two-orbital model with degenerate orbitals. Fig.~\ref{fig:2orb_dmft} shows the spectral function at interaction strength $U = 1.6D$, $J = U/4$ for both half-filling and hole-doped cases ($n\approx 1.5, 1.7$).
At half-filling, our results reproduce the spectral shape reported in Ref.~\onlinecite{Ganahl_mps_2015} while providing a more accurate description of the low-energy features around $\omega/D=0$ compared to real-time simulation with the same bond dimension (see inset).
Furthermore, as the system is doped away from half-filling, we observe a smooth evolution of the spectral function, demonstrating that our method is stable across a range of doping levels in multi-orbital simulations.

\begin{figure}[tbh]
    \centering
    \includegraphics[scale=1]{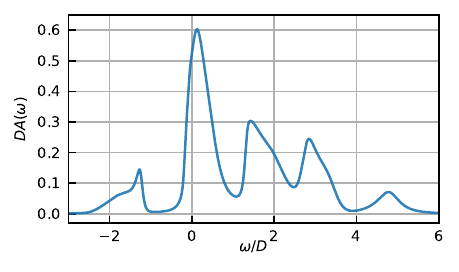}
    \caption{DMFT spectral function of the three-orbital Hubbard model with $D=1$, $U/D = 4$, $J = 0.15U$, $\mu/D=-1.0$ ($n \approx 1.0$). The complex-time simulation was performed with $\alpha=0.2$, $N_b = 239$, $Dt_\text{max}=60$, and $\chi=90$.}
    \label{fig:3orb_dmft}
\end{figure}

Fig.~\ref{fig:3orb_dmft} presents the spectral function for a three-orbital model with parameters inspired by those of vanadate and ruthenate transition metal compounds~\cite{Werner_spin_2008,Medici_Janus_2011,Bauernfeind_ftps_2017}. 
We set the half-bandwidth to $D=1$, the interaction strengths to $U/D = 4$, $J = 0.15U$, and the filling to roughly one electron per site, while employing a simplified different hybridization function [cf. Eq.~\eqref{eq:hybridization}]. 
As is evident from the figure, our complex-time plus ESPRIT-based analytic continuation approach clearly resolves the characteristic multiplets structure -- hole excitation, high-spin state, low-spin state, and Hubbard band -- associated with the Kanamori interaction, which is difficult to capture in imaginary-time simulations~\cite{Bauernfeind_ftps_2017}.
The resulting peak positions and heights are comparable to those reported for $\text{SrVO}_3$ in Ref.~\onlinecite{Bauernfeind_ftps_2017} while we do not introduce the additional broadening procedure to obtain smooth spectral functions near the Fermi level.

\section{Conclusions and Discussions} \label{sec:conclusion}
 
In this manuscript, we integrate a complex-time tensor-network impurity solver with an ESPRIT-based analytic continuation scheme for real-frequency DMFT calculations.
Using single-orbital impurity benchmarks, we demonstrate that this method reliably reconstructs real-time results from complex-time simulations, yielding accurate spectral functions and self-energies even for data obtained at relatively large complex angles.
The choice of complex angle $\alpha$ involves a balance between accuracy and efficiency. Larger $\alpha$ requires a smaller bond dimension to converge by suppressing entanglement growth, and the simulation time decreases with increasing $\alpha$ even at a fixed bond dimension \cite{cao_complextime_2024}. 
At the same time, increasing $\alpha$ will also make recovering spectral information more difficult.
We then show that the resulting method is both stable and computationally practical within zero-temperature real-frequency DMFT for both single- and multi-orbital systems.

These successful benchmarks, together with the improved computational efficiency, show that our method addresses a need for reliable and fast multi-orbital quantum impurity solvers that can access spectral information.
Such solvers are particularly important in the study of multi-orbital phenomena driven by Hund's coupling \cite{Werner_spin_2008, Georges_Hundreview_2013} such as the "Janus-faced" behavior \cite{Werner_spin_2008,Medici_Janus_2011} and the orbital-selective Mott transition \cite{Vojta_osmt_2010, Koga_osmt_2004, Arita_osmt_2005, Medici_osmt_2005}.
Overall, by enabling high-fidelity spectral resolution across a wide parameter regime, our method offers a promising pathway for {\it ab initio} studies of strongly correlated materials. 

\begin{acknowledgments}
X. Cao acknowledges support from the National Key R\&D Program of China under grant No.2024YFA1408602. 
X. Dong acknowledges support from the National Natural Science Foundation of China under grant No.12504289.
Y. Yu was supported by the National Science Foundation under Grant No. NSF DMR 2401159. 
E. Gull acknowledges funding from the European Research Council (ERC) under Advanced Grant No. 101142136 (Quantum Algorithms). 

\end{acknowledgments}

\appendix

\section{Non-interacting benchmark}\label{app:nonint}

\begin{figure}[tbh]
    \centering
    \includegraphics[scale=1]{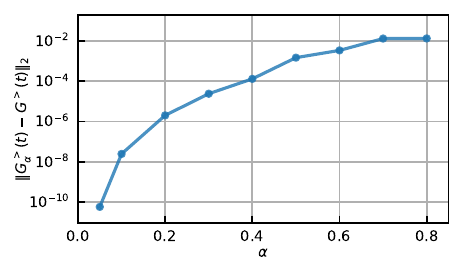}
    \caption{Error in the reconstructed greater Green's function as a function of the complex angle $\alpha$.}
    \label{fig:err_angle}
\end{figure}
%
\begin{figure}[tbh]
    \centering
    \includegraphics[scale=1]{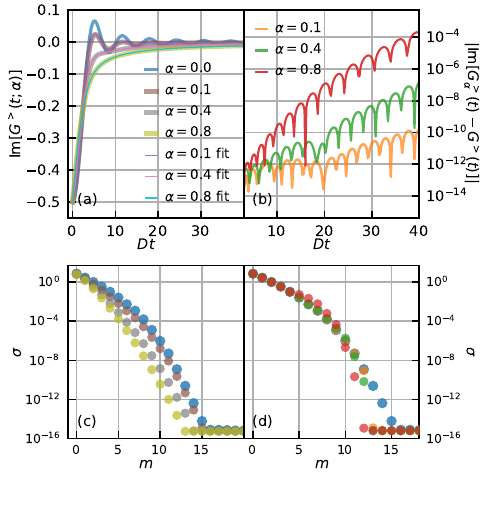}
    \caption{(a) Greater Green's function computed along complex-time contours with different angles $\alpha$. (b) Error in the corresponding reconstructed real-time Green's functions. (c)--(d) Singular value spectra of the Hankel matrices derived from the complex-time and the real-time Green's function in (a) and (b), respectively. }
    \label{fig:semi_angle}
\end{figure}

To analyze the performance of the ESPRIT-based analytic continuation, we use a single-orbital non-interacting impurity problem as benchmark, where a semielliptic density of states with half-bandwidth $D=2$ defined on $[-2.5, 2.5]$ is discretized into 1001 bath sites.
The real- and complex-time series are computed by diagonalizing the Hamiltonian and propagating with a time interval of $D\Delta t=0.2$ up to $Dt_\text{max}=60$. The error level is set to $10^{-15}$ in all ESPRIT fitting.

Fig.~\ref{fig:err_angle} shows the error in the greater Green's function $G_{\alpha}^{>}(t)$ reconstructed from simulated data at different complex angles $\alpha$. 
As is evident from the figure, the error increases with $\alpha$.
This increase in error originates from the loss of information at larger angles, as analyzed in Fig.~\ref{fig:semi_angle}. 
While the ESPRIT algorithm accurately fits the data for all angles [Fig.~\ref{fig:semi_angle}(a)], the singular value spectrum of the corresponding Hankel matrix reveals an important distinction [Fig.~\ref{fig:semi_angle}(c)]: the singular values decay faster at larger angles and fewer singular values are above the noise floor (i.e., machine precision in this example). In singular spectrum analysis, such behavior indicates that the signal is governed by fewer dominant physical modes. Consequently, the signal subspace obtained at large angles may not contain sufficient information to resolve fine details in the real-time domain.
This limitation is directly visible in the reconstructed real-time data [Fig.~\ref{fig:semi_angle}(b)], where the long-time behavior at larger angles shows significantly reduced accuracy.  Correspondingly, fewer singular values of the recovered real-time signal match with the ground truth [Fig.~\ref{fig:semi_angle}(d)].

\section{MaxEnt analytic continuation} \label{app:maxent}

\begin{figure}[tbh]
    \centering
    \includegraphics[scale=1]{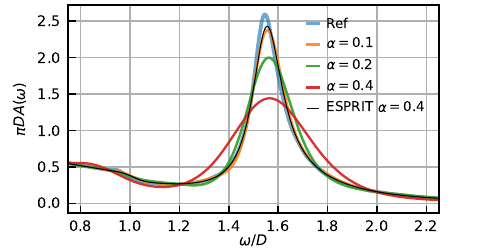}
    \caption{Comparison of spectral functions obtained from MaxEnt (colored lines) and ESPRIT (black line) with different complex angles at $U=2D$, $N_b=199$, $D t_\text{max}=80$ and $\chi=80$. The reference result is computed with $\alpha=0$, $D t_\text{max}=100$ and $\chi=1500$.}
    \label{fig:maxent_comparsion}
\end{figure}

For comparison, we analytically continue the complex-time results at various angles using the MaxEnt method as detailed in Ref.~\onlinecite{Guther_maxent_2018}. 
The relation between the complex-time Green's function and the spectral function is given by
\begin{align}
    G^{>}(z_{1}) &= - \mathrm{i} \int_{-\infty}^{\infty} \mathrm{d}\omega  \mathrm{e}^{-\mathrm{i} \omega z_{1}(t)} \theta(\omega) A(\omega), \\
    G^{<}(z_{2}) &= \mathrm{i} \int_{-\infty}^{\infty} \mathrm{d}\omega  \mathrm{e}^{-\mathrm{i} \omega z_{2}(t)} \theta(-\omega) A(\omega). 
\end{align}
This approach modifies the standard MaxEnt procedure by treating the greater and lesser components separately and generalizing the kernel $K(t,\omega)=\exp(-\mathrm{i} \omega t)$ to $K[z(t),\omega]=\exp[-\mathrm{i}\omega z(t)]$, where $z(t)=z_{1}(t)$ for $\omega\geq 0$ and $z(t)=z_{2}(t)$ for $\omega <0$. 
We employ a flat default model and determine the scaling hyperparameter for the entropy term using {\it Bryan's method}~\cite{Bryan_maxent_1990}. 
While MaxEnt more or less resolves the main features of the spectral function, it requires additional smoothing near $\omega=0$ as detailed in Ref.~\cite{grundner_complextime_2024} and also suffers from systematic amplitude decay at high frequencies as $\alpha$ increases (colored lines in Fig.~\ref{fig:maxent_comparsion}).
In contrast, the proposed ESPRIT method addresses this problem in a different way. It is able to accurately captures low-frequency behavior at various $\alpha$ (c.f. Fig.~\ref{fig:spectral_imp}) and preserves high-frequency spectral weight to a relatively good extent even at a relatively large $\alpha$, as shown by the black line in Fig.~\ref{fig:maxent_comparsion}.

\bibliographystyle{apsrev4-1}
\bibliography{ref}
\end{document}